\documentclass[aip,pop,reprint,amsmath,amssymb,longbibliography]{revtex4-2}
\usepackage{color}
\usepackage{mathrsfs}
\usepackage{graphicx}
\usepackage{dcolumn}
\usepackage{bm}
\usepackage{empheq}
\usepackage[hidelinks]{hyperref}
\usepackage{xcolor}
\hypersetup{
    colorlinks,
    linkcolor={blue},
    citecolor={blue!80!black},
    urlcolor={blue!80!black}
}
\usepackage[toc,page]{appendix}
\usepackage[normalem]{ulem}
\usepackage{amsmath,amsfonts,amssymb,ulem}
\usepackage{epstopdf}
\usepackage{xcolor}
\usepackage[separate-uncertainty]{siunitx}
\usepackage{upgreek}
\usepackage{hyperref} 

\usepackage{graphicx}

\makeatletter
\def\@fnsymbol#1{\ensuremath{\ifcase#1\or a)\or b)\or c)\or d)\else\@ctrerr\fi}}
\makeatother

\begin{document}

\title{Laser wakefield acceleration and x-ray emission in plasma produced by the electrical explosion of an aluminum wire}

\author{Heng Wang}
\affiliation{ 
Department of Engineering Physics, Tsinghua University, Beijing 100084, China
}%

\author{Liang Sheng}%
 \email{shengliang@nint.ac.cn}

\author{Jun Liu}
 \email{liujun@nint.ac.cn}

\author{Yang Li}

\affiliation{%
National Key Laboratory of Intense Pulsed Radiation Simulation and Effect, Northwest Institute of Nuclear Technology, Xi' an 710024, China
}%

\author{Yanfei Li}

\affiliation{%
Department of Nuclear Science and Technology, School of Energy and Power Engineering, Xi' an Jiaotong University,
 Xi' an 710049, China
}%

\author{Xinjian Tan}

\author{Dongwei Hei}

\affiliation{%
National Key Laboratory of Intense Pulsed Radiation Simulation and Effect, Northwest Institute of Nuclear Technology, Xi' an 710024, China
}%

\date{\today}
\begin{abstract}

Underdense plasmas used in relativistic laser-plasma interactions are conventionally produced from gaseous targets, whereas low-density metallic plasmas remain comparatively unexplored. Here, we investigate the interaction of a femtosecond petawatt laser with expanded aluminum plasma using particle-in-cell simulations based on an experimentally motivated exploding-wire density profile. Three-dimensional simulations show that the laser propagates through the low-density metallic plasma, drives a pronounced wake-like structure, accelerates electrons to near-GeV energies, and produces directional broadband x-ray emission. Charge-state-resolved diagnostics further show that electrons associated with higher aluminum ionization stages are disproportionately represented in the energetic population. Complementary two-dimensional simulations indicate that the net electron energy gain is dominated by longitudinal electric-field work and show that relativistic electron generation persists under variations of the assumed initial ionization and density conditions, although the yield remains quantitatively sensitive to the plasma state. These results demonstrate the feasibility of electrically exploded aluminum plasma as a target platform for laser wakefield acceleration and associated x-ray generation.

\end{abstract}
\maketitle

\section{INTRODUCTION}

The interaction of ultrashort, relativistically intense laser pulses with underdense plasmas provides a compact route to high-gradient electron acceleration and ultrafast radiation generation \cite{Tajima1979_19,Esarey2009Physics,Malka2012_2}. In laser wakefield acceleration (LWFA), the ponderomotive force of an intense laser pulse drives a large-amplitude plasma wave whose longitudinal electric field can trap and accelerate electrons to relativistic energies over millimeter-to-centimeter distances \cite{Esarey2009Physics,Malka2012_2}. The early experimental demonstration of quasi-monoenergetic relativistic electron beams \cite{Mangles2004_20,Geddes2004_8,Faure2004_21} has subsequently been extended to multi-GeV energies through improved laser guiding and plasma control \cite{Leemans2014_23,Gonsalves2019_9}. Accelerated electrons can also undergo transverse betatron oscillations in the focusing fields of the wake and emit broadband, forward-directed synchrotron-like radiation on femtosecond time scales \cite{Rousse2004Production,Kostyukov2003Xray,Corde2013Femtosecond,Kneip2010_22}. These developments have established LWFA as a promising basis for compact relativistic electron and x-ray sources and have made control of the plasma medium, in addition to control of the driving laser, a central aspect of accelerator design.

Gas jets, gas cells, and preformed plasma channels remain the most widely used media for LWFA because they provide controllable underdense plasmas over extended interaction lengths \cite{Esarey2009Physics,Malka2012_2,Audet2018_5}. The plasma density profile itself has increasingly become an active control parameter: sharp density transitions can modify electron injection \cite{Schmid2010_6}, longitudinal density tailoring can alter the injection and acceleration process \cite{Gonsalves2011_7}, and preformed plasma channels or capillary-discharge waveguides can guide intense laser pulses beyond the vacuum Rayleigh length \cite{Geddes2004_8,Leemans2014_23,Gonsalves2019_9}. These developments motivate broader approaches in which the plasma state is engineered before the relativistic pulse arrives. Solid materials can, for example, be converted into low-density plasma through laser ablation: laser-produced plasma plumes have been used for wakefield acceleration \cite{Rao2013_10}, and aluminum plasmas generated from solid targets have been investigated as an LWFA medium \cite{Kim2015_11,Kim2016_12}. Recent numerical studies have also explored hybrid targets in which overdense or nanostructured components are combined with adjacent underdense plasmas to control electron injection or radiation generation \cite{Cai2026LaserDriven,Caetano2026Electron}. Nevertheless, low-density plasmas derived directly from electrically prepared metallic targets remain much less explored than conventional gaseous media.

Metallic wires provide a useful system for distinguishing direct solid-target interaction from interaction with a deliberately prepared low-density metallic plasma. When an intense femtosecond laser directly irradiates a solid-density wire, experiments have demonstrated the generation and axial transport of energetic electrons \cite{Tokita2011_13,Nakajima2013_14}, together with strong terahertz surface electromagnetic waves propagating along the wire \cite{Tokita2015_15,Shao2025_16}. In such configurations, the target remains predominantly overdense and the interaction is governed largely by laser coupling near the solid surface, energetic-electron transport, transient charge separation, and surface fields. Electrical explosion provides a distinct preparation route. A high-current electrical pulse deposits energy resistively into the wire, driving rapid heating, phase transitions, ionization, and radial expansion \cite{Oreshkin2020Wire}. Experiments and magnetohydrodynamic modeling of aluminum-wire explosions have shown that the expanding material can separate into a relatively dense core and a rapidly expanding lower-density corona \cite{Sarkisov2005Nanosecond}. The energy deposition and subsequent expansion depend on the driving-current conditions and wire properties, providing experimentally accessible means of modifying the evolving plasma state \cite{Shi2015Experimental,Oreshkin2020Wire}. Electrical explosion can therefore transform an initially solid metallic wire into an extended, partially ionized plasma before the arrival of the relativistic laser. Earlier work considered laser heating and x-ray emission from exploding-wire plasmas in dense, collisional regimes \cite{Duston1978_18}; however, whether a realistic expanded metallic plasma produced by electrical explosion can support relativistic laser propagation, wakefield electron acceleration, and associated x-ray emission has received little attention.

Here, we investigate this possibility using an aluminum atomic-density distribution derived from the experimentally and computationally characterized exploding-wire conditions reported by Sarkisov \textit{et al.} \cite{Sarkisov2005Nanosecond}. The electrical explosion and the femtosecond laser interaction are treated as sequential stages: the former establishes the initial atomic-density distribution, whereas the subsequent relativistic pulse modifies the aluminum charge-state distribution through field ionization. Three-dimensional ionizing PIC simulations are used to examine laser propagation, wake formation, energetic-electron generation, photon emission, and the charge-state origin of the accelerated electrons. Complementary particle-resolved two-dimensional simulations are used to characterize the electron energy-transfer process and test the sensitivity of the energetic-electron population to variations in the assumed initial ionization and density conditions. The simulations show that expanded aluminum-wire plasma can support relativistic laser propagation, pronounced wake excitation, near-GeV electron acceleration, and directional broadband x-ray emission. Electrons associated with several higher ionization stages are disproportionately represented in the energetic population, whereas the principal acceleration behavior persists over the range of initial plasma conditions examined. These results establish electrically exploded aluminum plasma as a physically realizable target platform for relativistic wakefield acceleration and associated x-ray generation.

\section{PHYSICAL MODEL AND SIMULATION SETUP}

\begin{figure*}
\centering
\includegraphics[width=0.80\textwidth]{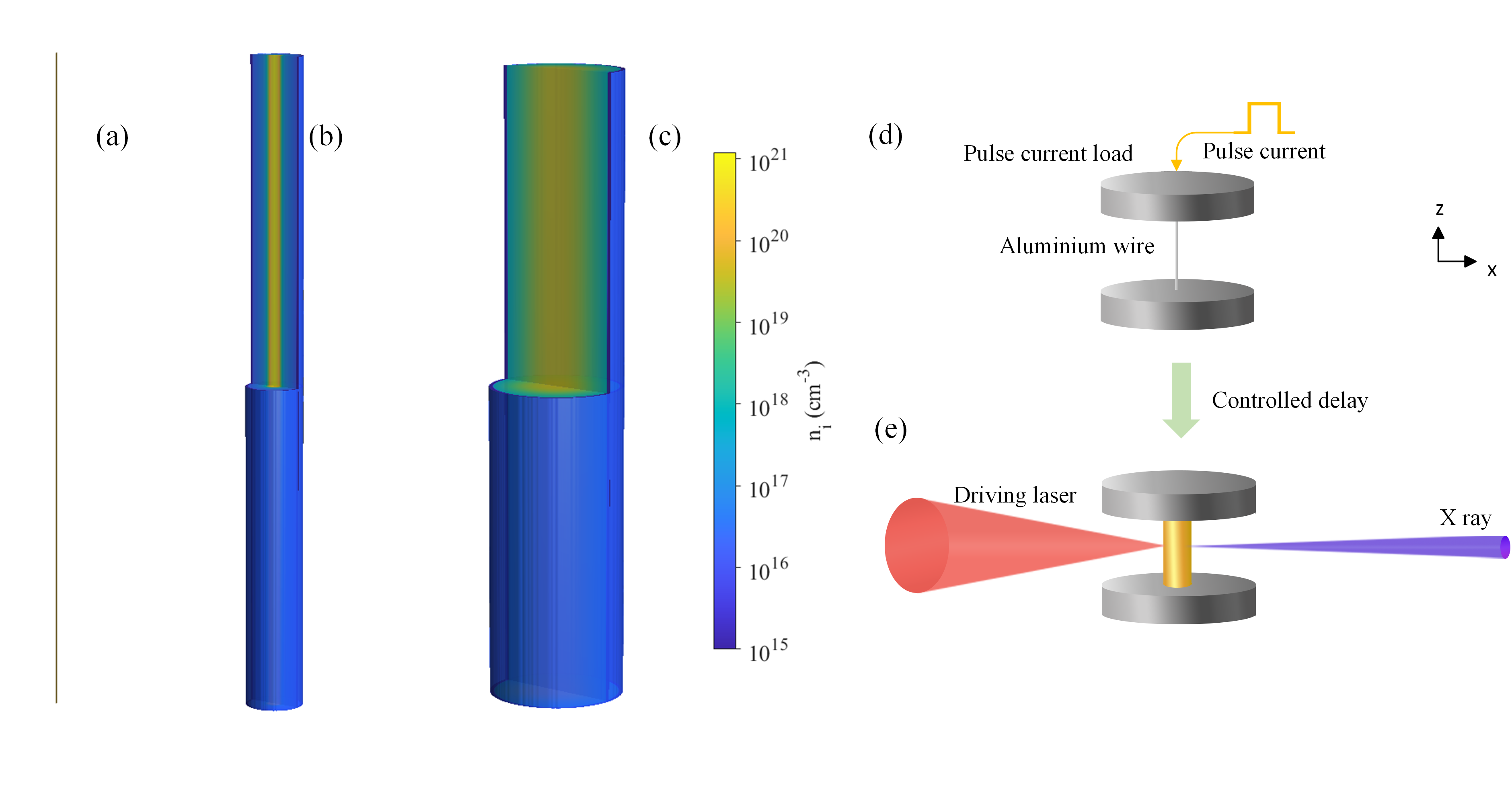}
\caption{Evolution of the aluminum atomic-density distribution used to construct the exploded-wire target: (a) initial wire, (b) 30~ns, and (c) 160~ns after the onset of the electrical pulse. (d) Schematic of the electrical-explosion stage. (e) After a controlled delay, a relativistic femtosecond laser irradiates the expanded plasma and produces energetic electrons and associated x-ray emission.}
\label{fig:configuration}
\end{figure*}

Figure~\ref{fig:configuration} illustrates the target concept and its implementation in the PIC simulations. The target is based on plasma produced by the electrical explosion of an aluminum wire. Rapid electrical energy deposition heats and ionizes the material and drives radial expansion. Experiments and magnetohydrodynamic modeling have shown that exploding aluminum wires can develop a relatively dense central region surrounded by an expanding lower-density plasma, with the evolution depending strongly on the current waveform and deposited energy \cite{Sarkisov2005Nanosecond,Shi2015Experimental,Oreshkin2020Wire}. The present scheme uses this expanded metallic plasma as the target for a subsequent relativistic femtosecond laser pulse [Figs.~\ref{fig:configuration}(d) and \ref{fig:configuration}(e)].

The atomic-density profile is constructed from the representative exploding-wire calculation reported in Ref.~\cite{Sarkisov2005Nanosecond} for a 15~$\mu$m-diameter aluminum wire driven by a pulsed current with a peak value of approximately 1.7~kA. The laser-interaction stage is initialized from the profile at 160~ns, when the wire material has expanded substantially. Cylindrical symmetry about the original wire axis is used to map the radial atomic-density profile into the simulation geometry. The laser propagates perpendicular to the original wire axis and is focused on the expanded low-density part of the target. For the reference ionization preparation described below, the corresponding initial free-electron density in the laser-interaction region is below $0.05n_c$. Because the electrical explosion evolves on a nanosecond time scale whereas the relativistic interaction occurs on a femtosecond-to-picosecond time scale, the exploded-wire profile is prescribed as the initial condition for the PIC calculation rather than simulating the electrical discharge and relativistic interaction self-consistently in a single calculation \cite{Sarkisov2005Nanosecond,Oreshkin2020Wire}.

Three-dimensional PIC simulations are performed with EPOCH \cite{Arber2015Contemporary}. The laser propagates along the $x$ direction and is incident perpendicular to the original wire axis, taken along $z$. The pulse has wavelength $\lambda_0=800$~nm, peak normalized vector potential $a_0=21.5$, focal spot radius $w_0=10~\mu$m, and pulse duration $\tau=30$~fs. The corresponding critical electron density is
\begin{equation}
 n_c=\frac{\epsilon_0m_e\omega_0^2}{e^2},
\end{equation}
where $\omega_0$ is the laser angular frequency. The three-dimensional computational window is $50\times40\times40~\mu\mathrm{m}^3$, resolved by $1250\times400\times400$ cells with 10 macroparticles per cell. A moving window is activated at $t=150$~fs to follow the laser over propagation distances longer than the instantaneous longitudinal box size.

The reference aluminum target is partially pre-ionized: 10\% of the aluminum ions are initialized as $\mathrm{Al}^{1+}$ and the remaining 90\% are neutral. One free electron is loaded for each initially singly ionized ion to maintain local charge neutrality. Further field ionization is calculated self-consistently using the ionization model implemented in EPOCH \cite{Arber2015Contemporary}. In the tunneling regime, the model uses the Ammosov--Delone--Krainov ionization rate together with the barrier-suppression treatment implemented in the code \cite{Ammosov1986Tunnel,Arber2015Contemporary}. The dynamic treatment therefore allows successive aluminum ionization stages to release electrons according to the instantaneous local field rather than assuming a fully ionized target at the beginning of the interaction.

Photon emission from relativistic electrons is included in the three-dimensional production simulation using the Monte Carlo synchrotron-emission model implemented in EPOCH \cite{Ridgers2014Modelling,Arber2015Contemporary}. Photon macroparticles above 1~keV are retained for diagnostics of photon number, energy, angular distribution, and spatial evolution. The reported photon number is obtained from the summed photon weights, and the laser-to-x-ray energy conversion efficiency is evaluated from the total energy carried by the retained photon population relative to the incident laser-pulse energy.

Complementary two-dimensional PIC calculations are used for particle tracking and sensitivity scans. They employ the same exploded-wire atomic-density prescription and the same reference pre-ionization model, but they are not treated as quantitative substitutes for the three-dimensional production calculation. In the two-dimensional reference configuration, the moving simulation window spans $50\times40~\mu\mathrm{m}^2$, resolved by $1250\times400$ cells with 10 macroparticles per cell. The laser wavelength is 800~nm, the total peak intensity is $1.0\times10^{21}$~W~cm$^{-2}$, the transverse Gaussian scale is 5~$\mu$m, and the pulse is applied over a 25~fs temporal interval. The moving window starts at 150~fs and propagates along $+x$ at $c$. The reduced dimensionality makes trajectory-resolved work analysis and systematic variation of the uncertain initial target state computationally tractable. Quantitative beam and photon yields are therefore taken from the three-dimensional calculation, whereas the two-dimensional simulations are used primarily to diagnose acceleration physics and relative sensitivity. Because dimensionality can modify laser and wake amplitudes and the resulting beam properties, the two-dimensional results are interpreted qualitatively and comparatively \cite{Vieira2008OneToOne}.

\section{RESULTS AND ANALYSIS}

\subsection{Laser propagation and electron acceleration}

\begin{figure}
\centering
\includegraphics[width=0.48\textwidth]{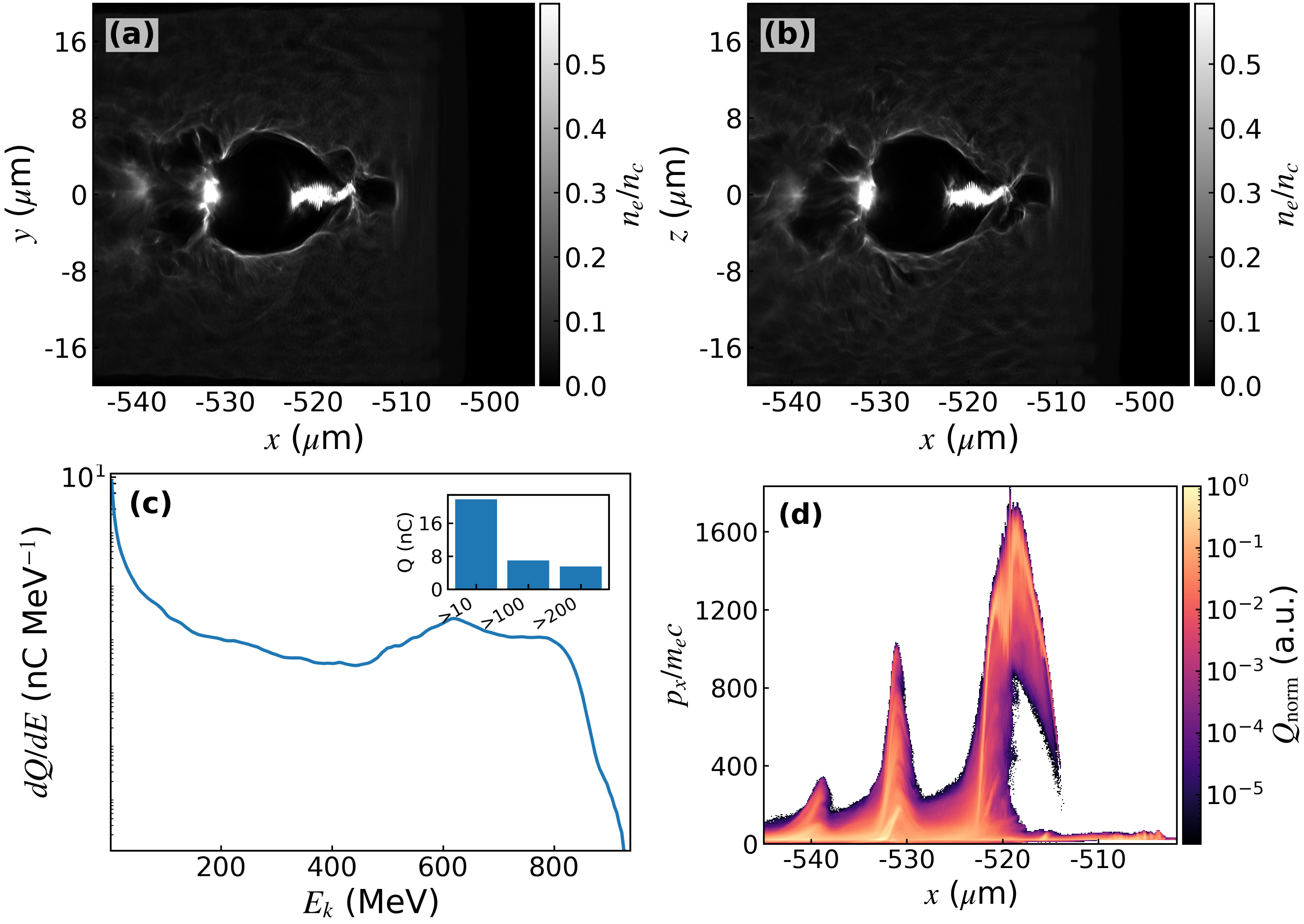}
\caption{Relativistic electron acceleration at $t=1$~ps in the three-dimensional PIC simulation. (a,b) Central $x$--$y$ and $x$--$z$ slices of the electron density normalized to the critical density, $n_e/n_c$, showing the channel and cavity structure. (c) Energy spectrum of forward-propagating electrons; the inset gives the integrated charge above representative energy thresholds. (d) Longitudinal $x$--$p_x$ phase space of electrons with $E_k>10$~MeV and $p_x>0$. The phase-space color scale denotes normalized particle-weighted charge density.}
\label{fig:accelerating}
\end{figure}

Figure~\ref{fig:accelerating} presents the relativistic laser interaction with the exploded aluminum plasma obtained from the three-dimensional simulation. Figures~\ref{fig:accelerating}(a) and \ref{fig:accelerating}(b) show electron-density distributions in the $x$--$y$ and $x$--$z$ planes, respectively. Despite the strongly nonuniform initial target profile, the laser propagates through the expanded plasma and produces a pronounced electron-depleted cavity followed by compressed electron structures. The morphology is characteristic of strongly nonlinear laser-driven plasma wakes \cite{Esarey2009Physics,Lu2006Nonlinear}. Similar structures in the two orthogonal planes show that the wake-like structure is genuinely three dimensional rather than an artifact of a single simulation plane.

The electron spectrum in Fig.~\ref{fig:accelerating}(c) contains a broad relativistic population with an energetic tail extending close to the GeV range. The integrated charge is of order tens of nanocoulombs above 10~MeV and several nanocoulombs above 100~MeV. The interaction therefore produces a substantial energetic-electron population rather than only a small number of isolated high-energy particles.

The longitudinal phase-space distribution in Fig.~\ref{fig:accelerating}(d) further shows that energetic electrons form localized populations behind the propagating laser pulse. Longitudinal momenta of order $10^3m_ec$ occur within the wake region. The spatial association between the wake structure and the high-$p_x$ populations is consistent with trapping and acceleration by a laser-driven plasma wake \cite{Esarey2009Physics,Lu2006Nonlinear}. Taken together, the three-dimensional results show that the expanded metallic target remains sufficiently transparent for relativistic laser propagation while supporting strong collective plasma fields and substantial electron acceleration.

\subsection{X-ray emission}

\begin{figure*}
\centering
\includegraphics[width=0.80\textwidth]{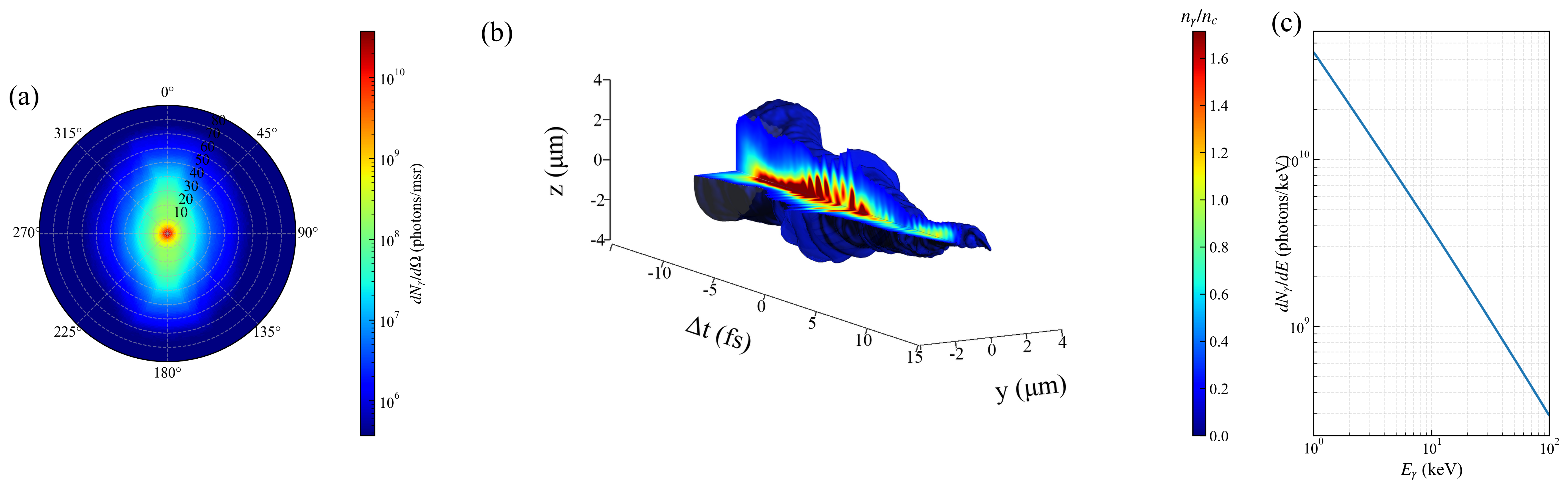}
\caption{X-ray emission from the three-dimensional PIC simulation. (a) Angular distribution of emitted photons. (b) Photon population plotted using the longitudinal diagnostic coordinate $\Delta t$ and transverse coordinates $y$ and $z$; the color scale gives photon number density normalized to $n_c$. (c) Photon-energy spectrum displayed from 1 to 100~keV.}
\label{fig:photon}
\end{figure*}

The accelerated electrons are accompanied by photon emission, as shown in Fig.~\ref{fig:photon}. Figure~\ref{fig:photon}(a) presents the angular distribution of emitted photons. The radiation is concentrated around the laser-propagation direction and decreases rapidly with increasing observation angle, demonstrating a pronounced forward-directed emission pattern.

Figure~\ref{fig:photon}(b) provides a complementary view of the photon population in the interaction region. The emission remains localized around the forward electron-propagation region and follows the evolving plasma channel. Such localization is consistent with radiation from relativistic electrons undergoing transverse motion in plasma focusing fields \cite{Kostyukov2003Xray,Rousse2004Production,Corde2013Femtosecond}.

At $t=1$~ps, the simulated photon population above the 1~keV diagnostic threshold reaches $2.39\times10^{11}$. The total photon energy corresponds to a laser-to-x-ray conversion efficiency of $3.39\times10^{-4}$. These integrated quantities are evaluated from the complete retained photon population; Fig.~\ref{fig:photon}(c) displays only the 1--100~keV portion of the spectrum. The displayed spectrum is broadband and decreases continuously through the plotted range, reaching photon energies of order $10^2$~keV. This broad, forward-directed emission is qualitatively consistent with betatron-like radiation from a relativistic electron ensemble with a distribution of energies and transverse oscillation amplitudes \cite{Kostyukov2003Xray,Corde2013Femtosecond}.

The three-dimensional simulation therefore demonstrates that electrically exploded aluminum plasma can support both relativistic electron acceleration and directional broadband x-ray emission within the same laser--plasma interaction. The radiation is treated here primarily as an associated output of the accelerated-electron dynamics rather than as an independently optimized x-ray source.

\subsection{Electron dynamics in the laser-driven wake}

\begin{figure*}
\centering
\includegraphics[width=0.80\textwidth]{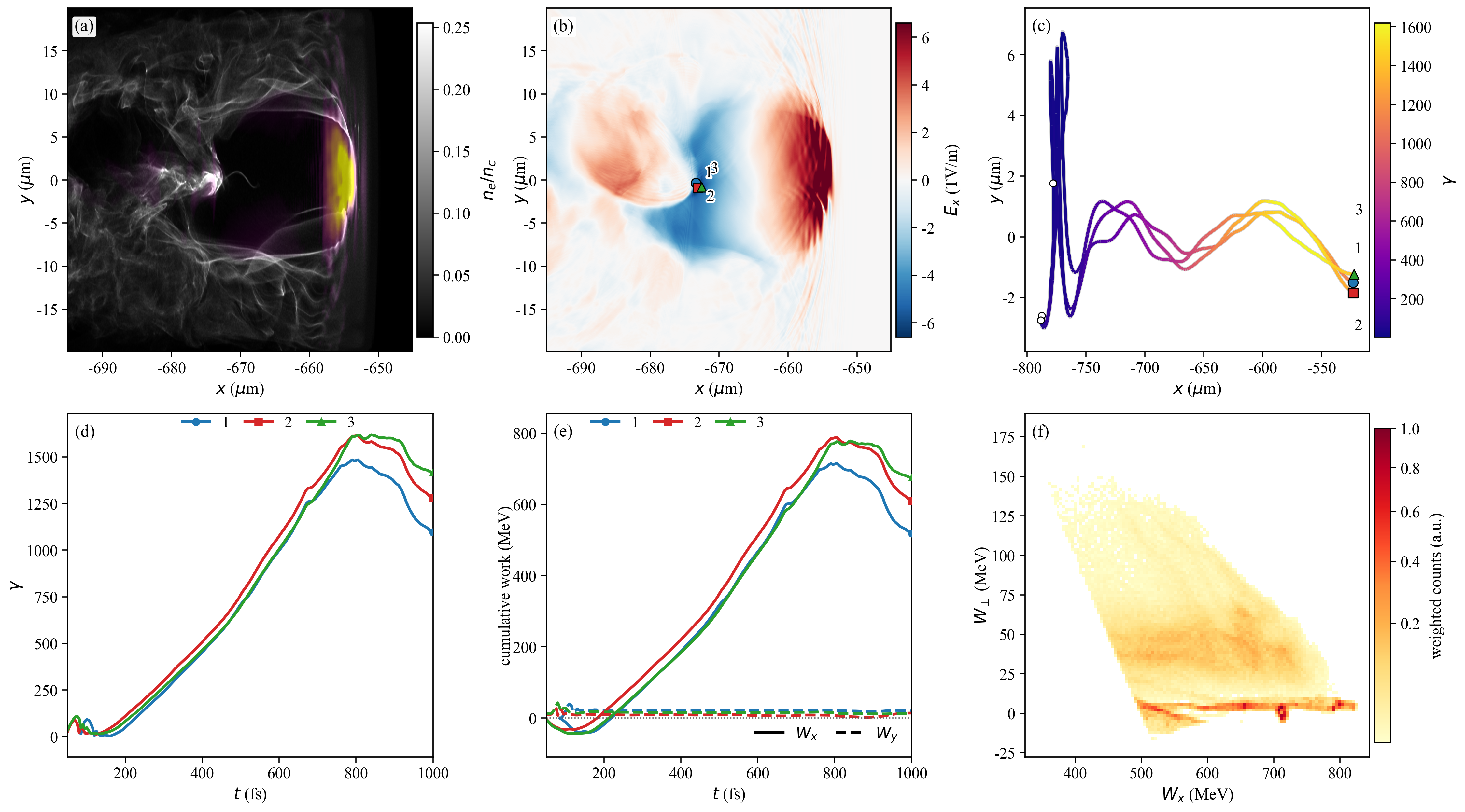}
\caption{Electron acceleration dynamics in the two-dimensional PIC simulation. (a) Electron density with the laser envelope overlaid at 500~fs. (b) Longitudinal electric field $E_x(x,y)$ at the same time, with three representative high-energy electrons marked. (c) $x$--$y$ trajectories of the same electrons, colored by $\gamma$. (d) Evolution of $\gamma(t)$. (e) Cumulative work histories for the representative electrons, with solid and dashed curves denoting $W_x$ and $W_y$, respectively; the dotted line marks $W=0$. (f) Weighted $W_x$--$W_\perp$ distribution for all final electrons with $E_k>500$~MeV, where $W_\perp=W_y+W_z$. The histogram is normalized and shown in arbitrary units.}
\label{fig:electron}
\end{figure*}

To characterize the acceleration process, particle-resolved diagnostics are performed using the two-dimensional simulation. The short laser pulse and wake-like structures in Figs.~\ref{fig:accelerating} and \ref{fig:electron} suggest wakefield acceleration, but trapped electrons can in some regimes acquire substantial additional energy directly from the transverse laser field when laser--electron overlap persists \cite{Shaw2017Role}. We therefore separate the work performed by the longitudinal and transverse electric fields.

Figure~\ref{fig:electron}(a) shows the electron-density distribution during the interaction. Consistent with the three-dimensional calculation, a pronounced low-density cavity and an extended plasma channel form behind the laser pulse. The corresponding longitudinal electric field $E_x$ is shown in Fig.~\ref{fig:electron}(b), where fields of several TV/m develop within and around the wake structure.

Three electrons that reach high final energies are selected for trajectory analysis. Their instantaneous positions are marked in Fig.~\ref{fig:electron}(b), and their $x$--$y$ trajectories are shown in Fig.~\ref{fig:electron}(c). The particles undergo appreciable transverse motion during the early interaction and subsequently propagate over an extended distance while their relativistic factors increase. Figure~\ref{fig:electron}(d) shows the corresponding temporal evolution of $\gamma$.

Because the magnetic field performs no work, the particle energy change can be decomposed into contributions from the electric-field components. For each tracked electron,
\begin{equation}
 W_i(t)=-e\int_{t_b}^{t}v_i(t')E_i(t')\,dt', \qquad i=x,y,z,
\end{equation}
where $t_b$ denotes the particle birth time for field-ionized electrons or the initialization time for preloaded electrons. We define the net transverse work as
\begin{equation}
 W_\perp=W_y+W_z.
\end{equation}
For the representative particles in Fig.~\ref{fig:electron}(e), $W_x$ rises to several hundred MeV and closely follows the energy evolution, whereas the displayed transverse contribution remains much smaller.

The same tendency is observed statistically in Fig.~\ref{fig:electron}(f). Most energetic electrons acquire substantially more net energy through the longitudinal field than through the total transverse field. The work decomposition therefore indicates that the energetic population is predominantly accelerated by the longitudinal plasma field under the conditions considered here. This behavior is consistent with wakefield-dominated acceleration \cite{Esarey2009Physics,Lu2006Nonlinear}, while allowing transverse fields to influence individual trajectories and oscillatory motion.

\subsection{Ionization-resolved electron acceleration}

\begin{figure*}
\centering
\includegraphics[width=0.80\textwidth]{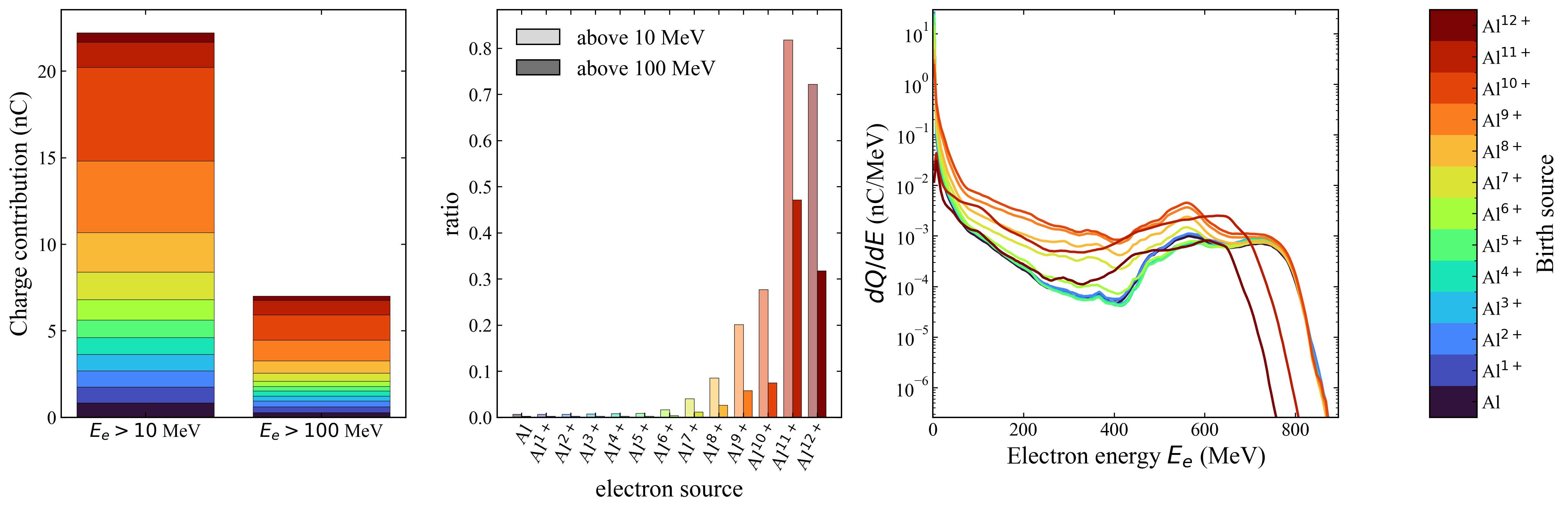}
\caption{Charge-state-resolved energetic-electron populations in the three-dimensional PIC simulation at $t=1$~ps. (a) Charge contributed by electrons associated with different aluminum source states for $E_e>10$ and 100~MeV. (b) Fraction of each source population reaching the corresponding energy threshold. (c) Electron energy spectra resolved by aluminum source state.}
\label{fig:ionization}
\end{figure*}

The aluminum target introduces sequential field ionization into the electron source. To determine whether electrons associated with different ionization stages contribute equally to the energetic population, the final electron distribution is resolved according to the aluminum source state recorded for each electron population in the three-dimensional simulation.

Figure~\ref{fig:ionization}(a) shows the charge contributions from the different source populations above 10 and 100~MeV. Electrons associated with a broad range of aluminum charge states contribute to the accelerated population, but the relative contribution from several higher stages becomes more pronounced for $E_e>100$~MeV.

This tendency is clearer when the energetic fraction within each source population is considered [Fig.~\ref{fig:ionization}(b)]. Several higher-stage populations have a substantially larger fraction of electrons above 10 and 100~MeV than lower-stage populations. Their enhanced representation in the energetic tail therefore cannot be explained solely by the number of electrons associated with each ionization stage. The charge-state-resolved spectra in Fig.~\ref{fig:ionization}(c) show the same behavior from an energy-resolved perspective: all source populations develop nonthermal tails, but their relative weights at high energy differ substantially, and the maximum energy does not vary monotonically with source charge state.

Charge-state-dependent electron release is known to influence injection in doped gaseous wakefield accelerators. Electrons with sufficiently large ionization potentials are released only where the local laser field is strong enough, changing their birth position and phase relative to electrons released earlier \cite{Pak2010Injection,McGuffey2010Ionization,Clayton2010SelfGuided}. The trend observed here is qualitatively reminiscent of this source-selection effect. The present charge-state-resolved spectra, however, do not by themselves establish that an identical ionization-injection mechanism operates in the expanded aluminum plasma. More conservatively, they demonstrate that the ionization stage associated with an electron is correlated with its probability of entering the energetic population. A complete reconstruction of the injection process would require simultaneous diagnostics of electron birth position, birth time, and subsequent wake phase.

\subsection{Sensitivity to the initial exploded-plasma conditions}

\begin{figure*}
\centering
\includegraphics[width=0.72\textwidth]{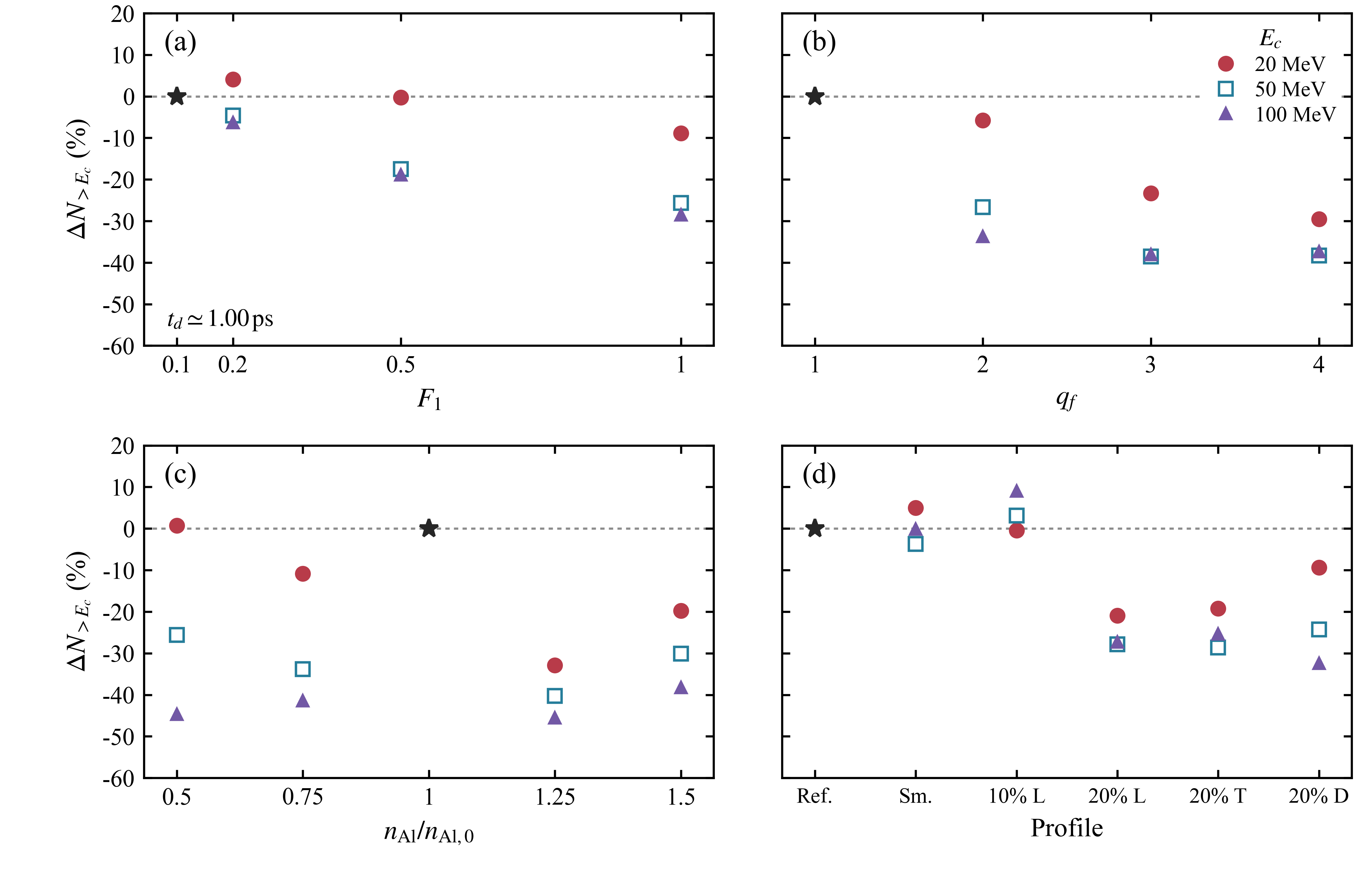}
\caption{Sensitivity of the energetic-electron population to the initial target conditions in the two-dimensional PIC simulations at $t_d\simeq1.0$~ps: (a) first-stage pre-ionization fraction $F_1$, (b) sequential initial charge-state preparation labeled by the frontier $q_f$, (c) aluminum atomic-density multiplier $n_{\mathrm{Al}}/n_{\mathrm{Al},0}$, and (d) representative density-profile modifications. The ordinate is $\Delta N_{>E_c}=100[N_{>E_c}/N_{>E_c,\mathrm{B1}}-1]$, where $N_{>E_c}$ is the sum of macroparticle weights of all electron species above $E_c$ within the moving simulation window. Circles, squares, and triangles denote $E_c=20$, 50, and 100~MeV, respectively; black stars mark the B1 reference. In (d), Ref. and Sm. denote the original and smoothed profiles, and L, T, and D denote prescribed longitudinal-, transverse-, and diagonal-dominant perturbations.}
\label{fig:sensitivity}
\end{figure*}

The state of an electrically exploded wire is not uniquely determined by the material alone. The deposited energy, current rise rate, and expansion history influence the density and ionization state encountered by the femtosecond pulse \cite{Sarkisov2005Nanosecond,Shi2015Experimental,Oreshkin2020Wire}. We therefore test whether the energetic-electron generation observed in the reference case relies on a narrowly specified initial target state.

Figure~\ref{fig:sensitivity} compares the electron populations above $E_c=20$, 50, and 100~MeV at the common diagnostic time $t_d\simeq1.0$~ps. The relative change is defined as
\begin{equation}
\Delta N_{>E_c}=100\left[\frac{N_{>E_c}(t_d)}{N_{>E_c,\mathrm{B1}}(t_d)}-1\right].
\end{equation}
Here, $N_{>E_c}$ is the summed macroparticle weight of all electron species above the specified energy threshold within the moving simulation window. It characterizes the energetic-electron population present in the computational domain at the diagnostic time and is not interpreted as cumulative injected charge or as a three-dimensional beam charge.

In the B1 reference, 10\% of the aluminum is initially $\mathrm{Al}^{1+}$ and the remaining 90\% is neutral. Figure~\ref{fig:sensitivity}(a) varies this first-stage pre-ionization fraction $F_1$. The population above 20~MeV changes moderately over much of the scan, whereas the populations above 50 and 100~MeV are more sensitive to the initial ionization state. Figure~\ref{fig:sensitivity}(b) considers related sequential charge-state preparations, with $q_f$ labeling the highest initially populated charge state. Increasing the degree of initial pre-ionization generally reduces the energetic-electron population, particularly at the higher thresholds.

The density scan in Fig.~\ref{fig:sensitivity}(c) varies the entire aluminum atomic-density profile between $0.5n_{\mathrm{Al},0}$ and $1.5n_{\mathrm{Al},0}$. Both lower and higher densities produce appreciable changes in the high-energy population, but energetic electrons remain present throughout the scanned interval. Figure~\ref{fig:sensitivity}(d) tests representative modifications of the spatial density profile, including smoothing of a narrow radial feature and prescribed 10--20\% perturbations with different dominant orientations. These changes produce variations from small enhancements to reductions of several tens of percent, but none of the tested profiles eliminates the energetic-electron population.

The sensitivity scans therefore distinguish quantitative variability from qualitative persistence. The high-energy yield depends on the exploded-plasma state, especially at the higher energy thresholds, but relativistic electron acceleration survives across the tested variations in initial ionization, density scale, and profile shape. This behavior is consistent with the broader experimental picture of electrical wire explosion, in which the plasma state evolves continuously with the discharge and expansion history rather than corresponding to a unique universal profile \cite{Sarkisov2005Nanosecond,Shi2015Experimental,Oreshkin2020Wire}.

\section{DISCUSSION}

Laser wakefield acceleration has conventionally been developed using underdense gaseous plasmas produced by gas jets, gas cells, capillary discharges, or related plasma-channel configurations \cite{Esarey2009Physics,Hooker2013Developments}. More recently, numerical studies have explored nonstandard and hybrid target architectures in which solid, nanostructured, or overdense regions are combined with underdense plasmas to modify electron injection or radiation generation \cite{Cai2026LaserDriven,Caetano2026Electron}. The target considered here differs from these approaches in that the low-density metallic plasma itself is produced before the relativistic interaction by electrical explosion of a metal wire.

The three-dimensional calculation demonstrates the central feasibility of this concept. A relativistic laser pulse propagates through the expanded aluminum plasma, drives a pronounced wake-like structure, and generates a substantial energetic-electron population together with directional broadband x-ray emission. These results are not intended to demonstrate superior accelerator or radiation performance relative to optimized gaseous LWFA or betatron sources. Rather, they show that an experimentally motivated class of expanded metallic targets can access a relativistic wakefield-acceleration regime normally associated with gaseous plasmas.

The particle-resolved two-dimensional diagnostics provide additional characterization of the acceleration. Previous work has shown that wakefield and direct-laser energy transfer can coexist when trapped electrons remain in contact with the driving laser \cite{Shaw2017Role}. In the present simulations, the cumulative longitudinal work substantially exceeds the transverse contribution for the energetic population examined in Fig.~\ref{fig:electron}. The principal net energy-transfer channel is therefore consistent with wakefield acceleration rather than a DLA-dominated regime \cite{Esarey2009Physics,Lu2006Nonlinear}. This analysis is used to identify the character of the acceleration and is not intended to establish a new acceleration mechanism.

The metallic composition nevertheless introduces an additional feature through sequential ionization. Ionization-controlled injection in gaseous LWFA depends strongly on the time and position at which more tightly bound electrons are released into the wake \cite{Pak2010Injection,McGuffey2010Ionization,Clayton2010SelfGuided}. The charge-state-resolved diagnostics in Fig.~\ref{fig:ionization} show an analogous selectivity in the expanded aluminum plasma: electrons associated with several higher ionization stages are disproportionately represented in the energetic population. At the same time, the sensitivity scans show that relativistic electron generation persists when the initial ionization preparation is varied. Dynamic ionization therefore influences the composition of the energetic-electron source without being the sole condition for the existence of the acceleration regime.

The accelerated electrons also produce a forward-directed broadband x-ray population. Its angular and spectral characteristics are qualitatively consistent with betatron-like emission associated with relativistic electrons undergoing transverse motion in plasma focusing structures \cite{Kostyukov2003Xray,Rousse2004Production,Corde2013Femtosecond}. The radiation has not been independently optimized here, and quantities such as source size, spectral brilliance, and tunability are not used as principal performance claims. The photon diagnostics instead establish that the proposed metallic-plasma platform can support an associated x-ray output within the same relativistic interaction.

The initial state of an electrically exploded wire introduces uncertainty absent from an idealized uniform plasma target. At the same time, electrical-explosion experiments show that deposited energy and expansion dynamics can be modified through the driving-current conditions \cite{Sarkisov2005Nanosecond,Shi2015Experimental,Oreshkin2020Wire}. The sensitivity results therefore have two implications. First, quantitative predictions of the highest-energy electron yield require knowledge of the plasma state at the time of laser arrival. Second, the same dependence suggests that the electrical pulse and laser delay may ultimately provide experimentally accessible control parameters for the relativistic interaction.

Several limitations should be emphasized. The nanosecond electrical explosion and the subsequent femtosecond laser interaction are not simulated self-consistently within one calculation; instead, a representative exploded-wire density profile is mapped into the PIC model. Binary collisional effects are also neglected in the relativistic interaction stage, which focuses on the strong-field sub-picosecond-to-picosecond dynamics after laser arrival. In addition, trajectory diagnostics and parameter scans are performed in two dimensions, whereas the principal electron and photon outputs and the charge-state-resolved source analysis are obtained from the three-dimensional simulation. Two-dimensional wakefield calculations can reproduce important qualitative trends while differing quantitatively from fully three-dimensional results \cite{Vieira2008OneToOne}. A more predictive treatment would ultimately couple a validated electrical-explosion model to three-dimensional relativistic PIC calculations over a wider range of experimentally accessible discharge and laser conditions.

\section{CONCLUSIONS}

We have investigated relativistic interaction of a femtosecond petawatt laser with expanded aluminum plasma produced by electrical wire explosion. An experimentally motivated exploded-wire density profile supports laser propagation through its low-density region, the formation of a pronounced wake-like structure, and acceleration of a substantial electron population to near-GeV energies in three-dimensional PIC simulations. The accelerated electrons are accompanied by forward-directed broadband x-ray emission, while charge-state-resolved diagnostics show that electrons associated with several higher aluminum ionization stages are preferentially represented in the energetic population. Two-dimensional particle tracking indicates that the net energy gain is dominated by longitudinal electric-field work, and sensitivity scans show that the acceleration persists under finite variations of initial ionization and density conditions even though the energetic-electron yield changes quantitatively. These results identify electrically exploded aluminum plasma as a feasible metallic target platform for laser wakefield acceleration and associated x-ray generation and motivate direct experiments combining pulsed-power target preparation with intense femtosecond laser irradiation.

\begin{acknowledgments}
This work was supported by the Project for the Advanced Pulsed Radiation Imaging Technology Research
Team of National Key Laboratory of Intense Pulsed
Radiation Simulation and Effect. 
Simulation results were obtained using the TH-eX system at the National Supercomputer Center in Tianjin. 
The authors gratefully acknowledge the EPOCH developer team for making their particle-in-cell code available, 
which was used for the simulations presented in this work.
The code EPOCH is funded by the United Kingdom Engineering and Physical Sciences Research Council Grants No. EP/G054950/1, No. EP/G056803/1, No. EP/G055165/1, and No. EP/ M022463/1. 

\end{acknowledgments}

\section*{Author Declaration}
\textbf{Conflict of Interest}: The authors have no conflicts to disclose.

\section*{Data Availability}
The data that supports the findings of this study are available from the corresponding author upon reasonable request.

\bibliography{aa}
\end{document}